\documentclass[%
 reprint,
 superscriptaddress,
 aps,
pra,
]{revtex4-2}

\usepackage{graphicx}% Include figure files
\usepackage{dcolumn}% Align table columns on decimal point
\usepackage{bm}% bold math
\usepackage{times}
\usepackage{float}
\usepackage{wrapfig}
\usepackage{xcolor}
\usepackage{graphicx}
\usepackage{graphics}
\usepackage{hyperref}
\usepackage[]{natbib}
\usepackage{physics}
\usepackage{qcircuit}
\usepackage{subfigure}
\usepackage{amsthm}
\usepackage{amsmath}
\usepackage{amssymb}
\usepackage{xcolor}
\usepackage[ruled]{algorithm2e}

\newcommand{\Ham}{\hat{\mathcal{H}}}
\newcommand{\X}[1]{\hat{\sigma}_{#1}^x}

\newcommand{\N}{\hat{n}}

\begin{document}

\title{Satellite Mission Planning with Rydberg Atoms}

\author{Michel Nowak}
\email{michel.nowak@thalesgroup.com}
\affiliation{Thales cortAIx Labs, Thales Research and Technology, Palaiseau, France}
\author{Benjamin Marchand}
\affiliation{Thales Alenia Space, Toulouse, France}
\author{Yassine Naghmouchi}
\affiliation{Pasqal, Palaiseau, France}
\author{Serge Rainjonneau}
\affiliation{Thales Alenia Space, Toulouse, France}
\author{Wesley Coelho}
\affiliation{Pasqal, Palaiseau, France}
\author{Louis Vignoli}
\affiliation{Pasqal, Palaiseau, France}
\author{Louis-Paul Henry}
\affiliation{Pasqal, Palaiseau, France}

\date{\today}% It is always \today, today,
             %  but any date may be explicitly specified

\begin{abstract}

Quantum computers relying on cold atoms are being built and promise a high flexibility in the way information in encoded into the physical system.
In particular, the analog mode is spiking interest in the field of optimization as a classically intractable number of configurations can be tackled.
In this work, we investigate a problem that requires every-day scheduling of critical tasks involving a large number of actors.
Namely, fixing the planning for a Earth Observation satellite fleet composed of several of units exposed to a high density of targets to be scanned.
We explore numerical schemes that convert the formulated problem into a cold-atoms friendly setup.
We begin by a naive formulation of the Satellite Mission Planning problem without taking account for the agility of the satellites.
We then extend the problem to take it into account based on the literature.
By formulating the planning problem as a Maximum Independent Set problem, we are able to solve the problem with a QPU based on Rydberg atoms.
We explore two ways of solving the MIS problem on the QPU, one relying on the graphs and on the Quadratic Unconstrained Binary Optimization Framework (QUBO).
We show that the QUBO methodology is the most relevant and explore it more deeply with numerical experiments.
We conclude on the potential utility of using a QPU to solve the Satellite Mission Planning problem in an operational context.

\end{abstract}

\maketitle

\section{Introduction}
\label{sec:introduction}
Optimization problems in an operational context are challenging to address for at least two reasons.
First, an optimal solution needs to be found within a reasonable time frame and with limited computational resources.
Time is often scarce, with constraints sometimes requiring a solution within a single day, within a single hour, or even less.
Second, determining an optimal strategy for solving an optimization problem incurs additional costs.
Metrics are necessary to assess how effective or inadequate a given strategy is.
In this work, we focus on optimizing resources for a satellite Earth observation mission.
Such a mission involves a satellite constellation managed by an operator and a set of observation requests to be performed on the following day.

In recent years, the capabilities and accessibility of quantum computers has significantly improved motivating research into their potential for efficiently solving problems that are computationally difficult on classical machines.
In particular, cold atom architectures~\cite{henriet2020quantum} can now support arrays of hundreds to thousands of atoms, and algorithms are being adapted to those platforms.

The aim of our work is to investigate whether a Quantum Processing Unit (QPU) based on Rydberg atoms can efficiently solve the Satellite Mission Planning problem. 
Our contributions are as follows:
\begin{itemize}
    \item Propose a strategy to encode the Satellite Mission Planning with Rydberg Atoms.
    \item Implement this approach using realistic scenarios.
    \item Provide a classical baseline for comparison.
    \item Compare performance metrics and solution times for both quantum and classical implementations.
    \item To our knowledge, this is the first use of a QPU based on Rydberg atoms to solve the Satellite Mission Planning problem.
    \item Release an open-source package to enable reproduction of our results.
\end{itemize}
The work is organized as follows:
Section~\ref{sec:scheduling_satellites} introduces the problem we focus on.
Section~\ref{sec:state_of_the_art} analyses the current approaches used in research and production.
Section~\ref{sec:planning_with_atoms} introduces the cold atoms architecture.
Section~\ref{sec:spot_library} introduces the python package we developed.
Section~\ref{sec:numerical_experiments} presents our approach.
Section~\ref{sec:conclusion} concludes the work.

\section{Satellite Mission Planning}
\label{sec:scheduling_satellites}

We begin by describing the problem we wish to solve, starting from the goals to reach, then enumerating the constraints that are imposed to the set of satellites.

\begin{figure}[h!]
    \centering
    \includegraphics[width=\linewidth]{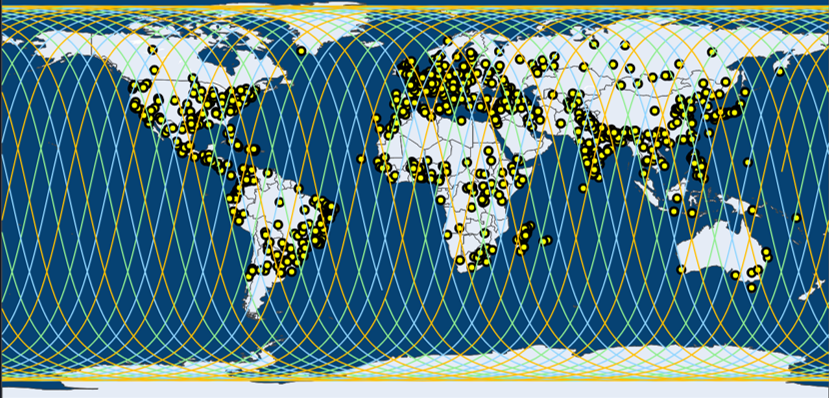}
    \caption{Visualization of requests of cities' observations and satellites' trajectories on earth. Yellow dots represent random cities on Earth. The trajectories are  Low Earth Orbit satellites. The goal of the scheduling task is to assign a request to each satellites.}
    \label{fig:visiting_cities}
\end{figure}

\subsection{Goals}
Each day, a list of observation requests—corresponding to zones of interest on Earth—is assigned to a constellation of satellites managed by a single scheduler.
The constellation can range from a single satellite to several hundreds. Figure~\ref{fig:visiting_cities} illustrates the problem setup.
Let $N_S$ represent the number of satellites available, and $N_R$ the number of observation requests to be completed within the next day. The scheduler's objective is to determine which satellite will carry out each request.

The primary performance metric is the total number of completed observation requests. The scheduler aims to maximize this metric while considering all flight and environmental constraints, which are detailed below.

\subsection{Constraints}
The scheduling of observation requests is subject to several constraints:
\begin{itemize}
    \item Power availability: The satellite is powered by a battery that recharges when exposed to sunlight. For simplicity, we assume that power consumption is linear with time and that the battery needs to be periodically recharged.
    \item Storage availability: The satellite has limited onboard memory. As observations are performed, memory is consumed and must be periodically freed by downloading data to a ground station. We assume a fixed amount of memory (M0) is used per observation, so the number of consecutive observations before each download is constrained by the ratio Mtot/M0, where Mtot is the total onboard memory.
    \item Maneuver preparation: Before each observation, the satellite must be reoriented towards the target area. Reorientation is performed at a limited speed to prevent adverse effects such as vibrations.
    \item Daylight operation: Optical satellites can only operate during daylight hours.
    \item Cloud coverage: Even if the satellite is positioned over the target area, it may not be able to complete the observation if the area is obscured by clouds.
   \item Rotation speed limit: The satellite can rotate about three axes to obtain a clear image of the target. Although the maximum speed is sufficient to ensure good image capture, rotation speed must be limited between acquisitions to avoid loss of image resolution.
    \item Target priority: Missions are assigned different priority levels based on the criticality of capturing the image. The scheduling algorithm must take these priorities into account, as the final metric depends on the number of missions completed at the highest priority levels.
    \item Operational flexibility: During daily operations, unplanned targets may need to be added en route. Handling such dynamic changes is an additional operational constraint for the scheduler.
\end{itemize}

Given the high number of constraints, some are often managed during pre- or post-processing. For example, cloud coverage can be incorporated into the Data Take Opportunity windows, while battery recharging can be assumed to occur when the satellite is passing over non-interesting zones and during daylight.

\subsection{Mathematical formulation}
Let $S$ be the set of satellites, referred to as $s$.
Let $T$ be the set of request target, referred to as $t$.
Let $W_{s,r}$ be the set of request target, referred to as $w$.

The parameters of the problem are:
\begin{itemize}
    \item $v_t$ value (priority) of target $t$.
    \item $c_{s,t,w}$ cost of satellite $s$ observing target $t$ in time window $w$.
    \item $C_s$ total cost budget for satellite $s$ (energy capacity)
    \item $m_{s,t,w}$ memory usage for satellite $s$ observing target $t$ in time window $w$.
    \item $M_s$ total onboard memory capacity for satellite $s$.
    \item $\Delta tt'$ slew (reorientation) time needed between two observations $t$ and $t'$ for satellite $s$.
\end{itemize}
The objective function is maximize the total value of observations
\begin{equation}
    \max\sum_{s\in S}\sum_{t\in T}\sum_{w\in w_{s,t}} v_t\cdot x_{s,t,w}\,
\end{equation}
where $x_{s,t,w}$ are binary decision variables.

The constraints are expressed as:
\begin{itemize}
    \item Each observation must be observed once
    \begin{equation}
        \sum_{t\in T}\sum_{w\in w_{s,t}} x_{s,t,w} \leq 1,\forall t \in T
    \end{equation}
    \item Energy consumption must be lower than satellite capacity
    \begin{equation}
        \sum_{t\in T}\sum_{w\in w_{s,t}}c_{s,t,w}\cdot x_{s,t,w} \leq C_s \forall s\in S
    \end{equation}
    \item Memory consumption must be lower than total memory onboard
    \begin{equation}
        \sum_{t\in T}\sum_{w\in w_{s,t}}m_{s,t,w}\cdot x_{s,t,w} \leq M_s \forall s\in S
    \end{equation}
    \item Agility constraint: for each pair of opportunity compute:
    \begin{equation}
        x_{s,t,w} + x_{s,t',w'} \leq 1 \text{ if } w'< w + \Delta tt'
    \end{equation}
\end{itemize}

\subsection{Metrics}
With these goals at hand, designing a solver that respects the large number of constraints remains a challenging task.
In order to guide the solver, we need to define high level metrics.
\begin{enumerate}
    \item \textbf{Global completion rate}: to be maximized
    \item \textbf{Satellite workload balance} to be equalized
    \item \textbf{Maneuver duration} to be minimized 
    \item \textbf{Image quality} quantified as distance to zero pitch (to be as close to $0$ as possible, but can have negative and positive values)  
    \item \textbf{Time to solution} to be minimized because operators of satellite constellations could iterate over the plans in order to better fit client's needs for example.
\end{enumerate}
\section{State of the art}
\label{sec:state_of_the_art}

\subsection{Classical approaches}
The problem of scheduling concurrent observations for a satellite fleet has been extensively covered with classical computers in~\cite{globus2004comparison}.
The problem is encoded in graphs and standard graph coloring algorithms are directly applied to obtain an approximate solution~\cite{marx2004graph}.
Classical algorithms are adapted if the problem is of too large size.
The graph coloring approach is adapted to large scheduling problems in~\cite{leighton1979graph}.
Another solution is to encode the problem into a Maximum Independent Set problem.
Here we review the strategy proposed by~\cite{eddy2021maximum}.
In order to tackle graphs of large size, one can resort to distributed MIS~\cite{ghaffari2016improved, bar2017distributed}.
There are many ways to decompose a big graph into subdomains and merge MIS~\cite{tatti2015density}.
Finally, an approach on graph neural networks is presented in ~\cite{brusca2023maximum, pontoizeau2021neural}.

Currently, heuristic scheduling techniques are widely used for the Agile Earth Observation Satellite Scheduling Problem which typically fail to provide bounds on the quality of the scheduling scheme.
\subsubsection{Exact Methods}
To address this issue, two Exact Methods, Branch and Bound (B\&B) and Mixed Integer Linear Programming (MILP) have been developed to provide optimal or near-optimal solutions.
Early work~\cite{gabrel1997new, gabrel2003mathematical} discretized visibility time windows (DTOs) and applied B\&B on directed graphs.
More recent work~\cite{valicka2019mixed} developed a deterministic MILP model offering optimality guarantees, later extended to handle stochastic scenarios like cloud uncertainty.
Other research, such as ~\cite{chu2017branch, cho2018optimization}, explored simplified linear assumptions to make MILP more tractable for real missions.

\subsubsection{Heuristics}
Heuristics offer faster, though potentially suboptimal, solutions. These include:
\begin{itemize}
   \item  \textbf{Constructive heuristics}, which build schedules step-by-step and are useful under complex operational constraints (for example~\cite{lemaitre2002selecting} work for the French Pleiades project). We can find methods like Greedy algorithm, Dynamic Programming method, Constraint Programming approach or Local Search algorithm.
   \item \textbf{Time-efficient heuristics} designed for autonomous satellites, addressing real-time challenges like cloud cover. These combine rule-based methods and reactive planning (for example autonomous framework~\cite{beaumet2008autonomous} and emergency task insertion strategies~\cite{song2018emergency}). Key methods include: Stochastic Greedy algorithm, Semi-Markov decision process and Monte Carlo Tree Search.
   
\end{itemize}

\subsubsection{Metaheuristics}
These general frameworks adapt to complex problems and dominate the Agile Earth Observation Satellite Scheduling Problem (AEOSSP). Key methods include:

\begin{itemize}
    \item  \textbf{Evolutionary Algorithms}, such as Genetic Algorithms (GA) and Ant Colony Optimization (ACO), which are popular for multi-objectives scheduling (e.g., profit maximization and fairness). For instance~\cite{li2018preference} combined GA with Simulated Annealing to enhance convergence speed and solution quality. ~\cite{yuan2014agile} introduced evolutionary strategies to rapidly find high-quality initial solutions, while ~\cite{du2018area} modeled the AEOSSP as a Traveling Salesman Problem and applied an ACO-based approach.
    
    \item \textbf{Single-point search algorithms}, like Tabu Search, Local Search, and Adaptive Large Neighborhood Search, are used for both single and multi-satellite scenarios, offering high-quality solutions with reasonable computational effort.
    Research like~\cite{habet2003saturated} treated the AEOSSP as a vehicle routing problem with time windows, and~\cite{cordeau2005maximizing} developed a simple yet effective Tabu Search strategy that earned second place in the ROADEF challenge competition~\cite{roadef2019challenge}.
    Adaptive large neighborhood search and iterated local search respectively, to explore complex scheduling landscapes efficiently can be found in ~\cite{liu2017adaptive, peng2018iterated}.
\end{itemize}

\subsubsection{Machine Learning}
Machine learning (ML) has growing potential in AEOSSP, especially for improving adaptability and autonomy in dynamic environments. Though currently limited in handling operational constraints, some promising approaches have emerged.
\begin{itemize}
    \item  \textbf{Deep learning-based scheduling}, like the Long Short-Term Memory (LSTM)-based model in~\cite{peng2018onboard}, enables fast decision-making by learning task features from historical data. Similarly, ~\cite{du2019data} proposed a data-driven parallel scheduling system using neuro-evolution to predict task priorities and enable distributed planning.
    \item \textbf{Reinforcement Learning} (RL), is also a natural fit for sequential decision-making problems like satellite scheduling. AEOSSP can be modeled as a Markov Decision Process (MDP) or Partially Observable MDP, allowing RL agents to learn optimal policies over time. For example, ~\cite{chun2023deep} considered AEOSSP as a Markov Decision Process (MDP) and employed a Graph Attention Network-based Decision Neural Network (GDNN) with a model trained using Proximal Policy Optimization (PPO).
    \item \textbf{Multi-agent systems} also show promise, especially in multi-satellite constellations where each satellite can act as an agent. These systems allow decentralized planning and real-time coordination. For example, ~\cite{marchand2025dynamic} demonstrates the advantages of addressing the Earth observation satellite scheduling problem using a self-Adaptive Multi-Agent System (AMAS) approach. These multi-agent systems could also  can be combined with learning-based techniques (like RL) for cooperative learning and task allocation.
\end{itemize}

\subsection{Quantum approaches}
Since the availability of quantum computing resources, studies have been carried out to get insights on the possibility of solving scheduling problems.

In the digital mode, the usage of Variational Quantum algorithms (VQA) along with a Rreinforcement strategy has been explored~\cite{rainjonneau2023quantum}.
A baseline is provided with inter programming.

In the analog mode, the first well defined satellite observation tasks scheduling was formulated with quantum annealing resources~\cite{stollenwerk2021agile} into a graph coloring problem.
The problem was tuned to fit to the compilation constraints of the D-Wave architecture.
Latest results can be found in~\cite{makarov2024quantum}.

\section{Planning with atoms}
\label{sec:planning_with_atoms}

We rely on the analog quantum computing mode of Rydberg atoms~\cite{henriet2020quantum}.
Rydberg architectures are now known to perform well on some of maximum independent set problems~\cite{eddy2021maximum}.
It can be understood easily. A graph is made of nodes and edges.
Each node is mapped to an atom.
Each edge is mapped to two atoms being close enough to activate the Rydberg blockade mechanism.
If you have a graph whose nodes correspond to possible outcomes of your optimization problem, then adding edges between incompatible nodes (constraints of your problem), then extracting as much atoms in the excited state is equivalent to solving the MIS problem on that graph.
Note that this strategy has never been applied on a quantum hardware, and this will be the first time in our project that we make atom-based architectures optimize satellite mission planning.

\subsubsection{Extracting Data Take Opportunities}
Let's explain how the MIS can be used to solve the mission planning problem.
Each satellite will have a different trajectory, which will allow it to take a set of certain set of observations. For each satellite, we define the Data Take Opportunity (DTO) as the time window during which a satellite can complete an requested observation.
One request might be present in more than one DTO per satellite, and one request might be performed by more than one satellite, which makes the problem computationally hard to solve.
\begin{figure}[h!]
    \centering
    \includegraphics[width=\linewidth]{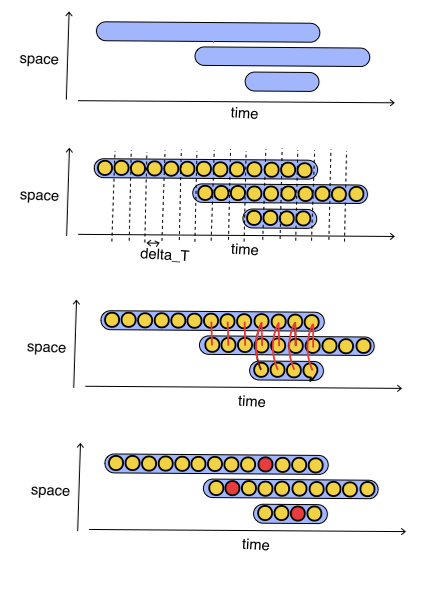}
    \caption{Formulation of the Maximum Independent Set problem with the EDGES method starting from data take opportunities. (a) One blue satellite can take multiple requests in space during the some random time windows (for illustration). (b) We split time during one day from $0$ seconds to $24\times24\times60$ seconds. Each yellow dot represent an opportunity time slot in which the observation can be made. It is mapped to an undirected graph (c) We add edges to this graph depending on the constraints. (d) we solve the Maximum Independent Set, thus retrieving independent and scheduoled opportunities for as many requests as possible.}
    \label{dtos_stepwise}
\end{figure}

A first research path taken is the one where satellites are assigned with a time slot for one request of observation. This is represented by a node containing (request\_id, time\_step, satellite\_id). We make the assumption that the maneuver time is included into the time slot.
For that, we chose time slots of one minute for each satellites, construct the graph of incompatibilities (between acquisition slots) and solve the Maximum Indendent Set problem on these graphs.
The incompatibilities are: A satellite cannot take 2 acquisitions in the same time step.
A request cannot be selected in the same time step by more than one satellite.
Figure~\ref{dtos_stepwise} illustrates 3 data take opportunities concurrent in time.
By selecting nodes in the independent set (as many as we can) will solve the problem and we will have extracted independent requests, so that all constraints are met.
\begin{figure}[h!]
    \centering
    \includegraphics[width=\linewidth]{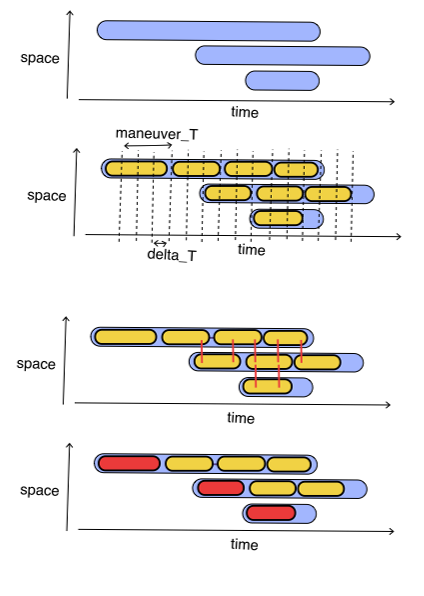}
    \caption{DTO generation with agility constraint}
    \caption{Formulation of the Maximum Independent Set problem starting with the AGILE method from data take opportunities. (a) One blue satellite can take multiple requests in space during the some random time windows (for illustration). (b) We split time during one day from $0$ seconds to $24\times24\times60$ seconds. Each yellow dot represent an opportunity time slot in which the observation can be made It is wider that with the edges method because we take into account the agility of the satellite. It is mapped to an undirected graph (c) We add edges to this graph depending on the constraints. (d) we solve the Maximum Independent Set, thus retrieving independent and scheduled opportunities for as many requests as possible.}
    \label{fig:dtos_agility}
\end{figure}

This procedure is valid, if the satellite moves quite fast compared to the 1 min time slot. It needs to perform both the maneuver and the acquisition in the defined time step.
However, in reality, the maneuver duration strongly depends on the attitude of the satellite (how it is oriented with pitch, roll, yaw, angles). This is why we add a maneuver routine which gives the time to fix a point on the globe. The time to maneuver can vary from 10 seconds to more than several minutes. A fixed speed on the apparatus is imposed to avoid vibrations and keep the quality of the image as good as possible through stabilization.
Figure~\ref{fig:dtos_agility} illustrate our new algorithm.
Each node is now generated by taking the time into account the maneuvre time, which reduces the number of nodes in the graph compared to the old method.

\begin{figure}[h!]
    \centering
    \includegraphics[width=0.48\linewidth]{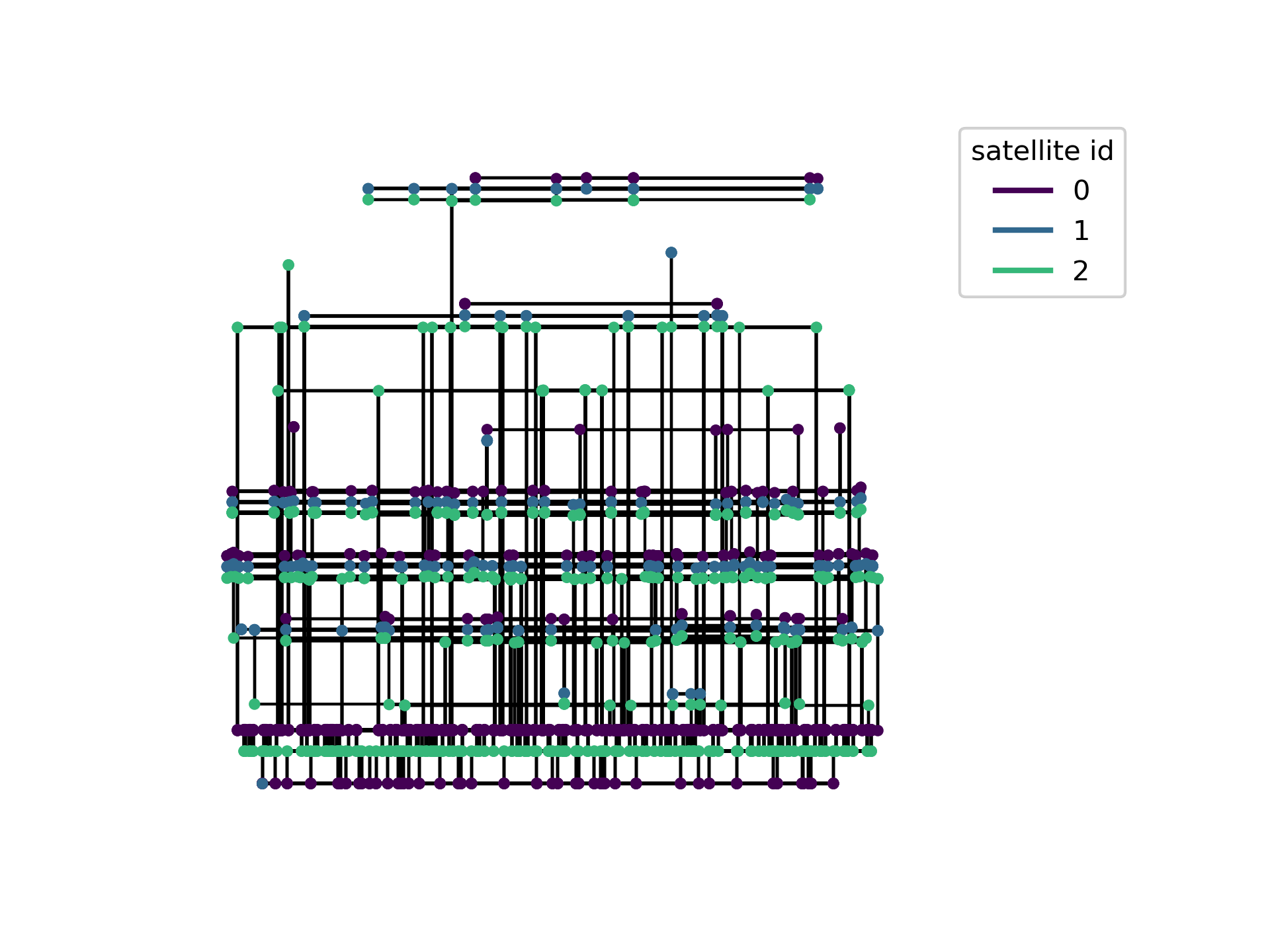}
    \includegraphics[width=0.48\linewidth]{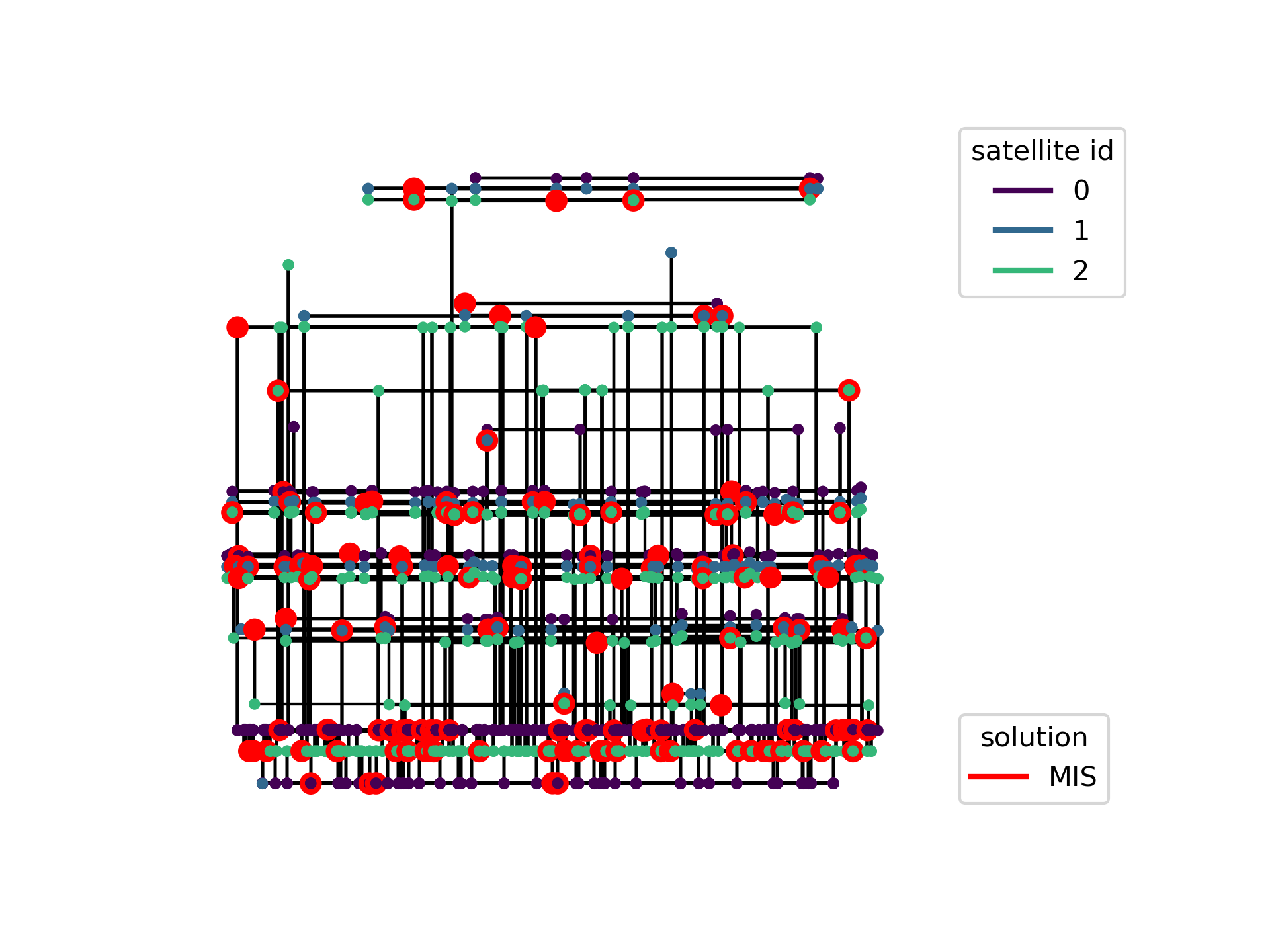}
    \caption{Left: solving graph whose MIS needs to be extracted to retrieve a relevant plan respecting the planning constraints. Right: with Maximum Independent Set solution.}
    \label{fig:dtos}
\end{figure}
Figure ~\ref{fig:dtos} show the resulting graphs for 3 satellites and few hundreds of acquisition requests during 1 day splitting into minutes.
Now, these 3D graphs are theoretical and cannot be used as is on the QPU directly.
Note that if some of the regions of the solving graph are disconnected, then this gives subgraphs of smaller size. The MIS on all subgraphs correspond to the MIS on the solving graph because disconnected  graphs are independent from each other.

The unit disk property of the graphs on the hardware needs to be preserved, which means that edges correspond to a distance lower than a fixed radius which is a hardware property.

\subsection{Autoencoders}
The first strategy we had to embed the graphs on a Unit Disk compatible to Rydberg atoms was to use autoencoders as proposed in~\cite{vercellino2022neural}.
The neural network encodes the positions of the atoms and minimizes a loss in which the constraints of the hardware are taken into account.
For example, two adjacent nodes will have an attracting loss, while unconnected nodes will be repulsively placed. The neural networks thus tries to reconstruct the positions of the atoms from an initial guess while having a penalty (regularization) for not respecting the constraints of the hardware.

\begin{figure}[ht!]
    \includegraphics[width=\linewidth]{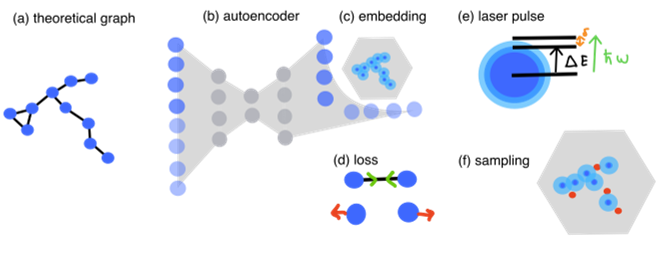}
    \caption{Autoencoder method~\cite{vercellino2022neural}. (a) The solving graph is computed and (b) embedded onto the QPU respecting the constraints (unit disk, Rydberg blockade, maximum distance between atoms). (d) The embedding is obtained by minimizing a loss which is tuned according to the edges of the graph with repulsive and attracting forces. (e) The atom is impinged with a detuned laser with a given amplitude during some range of time. (f) The experiment is measured and samples are retrieved. The MIS represent the possible tuples between a satellite, a time step, and a request.}
    \label{fig:pulser}
\end{figure}
Figure~\ref{fig:pulser} shows and example for 1 satellite. The initial graph is spitted into small disconnected graphs and each of these subgraphs are then embedded with an autoencoder, the positions are optimized and we can finally sample the MIS with the emulator of PASQAL's hardware.
In the middle of the figure, you can find the arranged atoms after optimization and mapping of their optimized position. On the right of the figure, you will find the excited nodes after measurement (emulation of hardware).
Although this method was implemented, we decided not to continue following it as it the training phase is too long compared to other embedding methods.

\subsection{QUBO on Rydberg atoms}

\subsubsection{Rydberg atoms reminders}

There are two ways to use Rydberg atoms.
The digital gate-based mode is not implemented yet in the literature, and we instead focus on the analog one, which dynamics is described by the Ising model with the following Hamiltonian.
\begin{equation}
    \Ham(t) =\hbar\sum_{i=1}^{N}  \left(\frac{\Omega(t) }{2} \X{i} - \delta(t)\,\N_i\right) + \sum_{i<j} \frac{C_6}{|\mathbf r_i - \mathbf r_j |^6} \N_i \N_j
    \label{eq:ising}
\end{equation}
where $\Omega$ is the amplitude of the laser that impinges the atoms, $\delta$ is the detuning with respect to the atomic resonance, $\mathbf{r}_i$ the position of a given atom $i$ and $C_6$ a constant that is associated with the type of atoms chosen, in our case Rubidium.

By tuning $\Omega$ and $\delta$, we can find a way to prepare the ground state of some Ising models. Note that only the spatial distribution needs to be tuned to take into account all of the degrees of freedom of the Hamiltonian. So how can this physical entity be used to solve an optimization problem? Let's look at the formulation of a generic QUBO model.

\subsubsection{QUBO and the Ising model}
A QUBO - standing for Quadratic Unconstrained Binary Optimization - can be written in the following form:
\begin{equation}
    g(x_1,\ldots,x_N) = \sum_{i>j} (Q_{ij} + Q_{ji}) x_i x_j + \sum_i Q_{ii} x_i
                  \equiv \mathbf x \; Q  \; {\mathbf x}^T,
\end{equation}
where $Q$ is the QUBO matrix containing coefficients.
If we set $\Omega$ to zero, we can recover this form for the energy:
\begin{equation}
    E(x_1,\ldots,x_N) = \sum_{i>j} \frac{C_6}{|\mathbf r_i - \mathbf r_j |^6} x_i x_j + \delta \sum_i x_i.
                    \equiv \mathbf x \; U  \; {\mathbf x}^T,
    \label{eq:spinglass}
\end{equation}
The off diagonal coefficients of $U$ correspond to couplings between on atom and another and the diagonal terms are the local detuning.
The analogy between QUBO and the Ising model is thus straightforward and minimizing the QUBO (which means solving the optimization problem) is thus equivalent to finding the fundamental of the corresponding Ising model.

\subsubsection{Minimization procedure}
In order to prepare such a fundamental state, we rely on a adiabatic methodology.
There are two hypothesis:
\begin{itemize}
    \item If we prepare the system in the fundamental state of the Hamiltonian at hand
    \item If we change the Hamiltonian sufficiently slowly compared to the energy gaps between the eigenmodes
\end{itemize}
then, we can maintain the system in the fundamental eigenmode of the Hamiltonian.
The initial state should be easily made:
\begin{equation}
  \ket{0}^{\otimes N}=\ket{0\ldots 0},
\end{equation}
For $\Omega=0$, we can write the initial Hamiltonian as:
\begin{equation}
    \Ham_0 = \hbar\sum_{i=1}^{N}  - \delta(0)\,\N_i + \sum_{i<j} \frac{C_6}{|\mathbf r_i - \mathbf r_j |^6} \N_i \N_j
\end{equation}
The constraints for this physical setup to be prepared are that the detuning needs to be positive($\delta$). Also, quantifying what "slow enough" means as expressed in the hypothesis remains challenging, and we fix the total evolution time as a hyper parameter of the experiment.
\begin{figure}[ht!]
    \includegraphics[width=\linewidth]{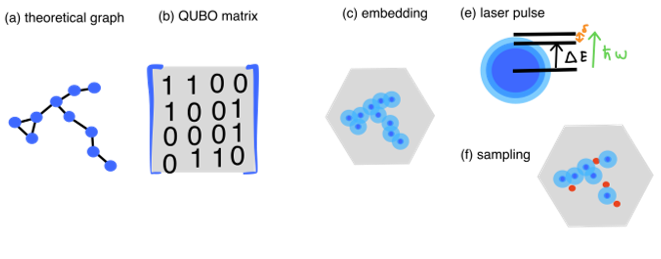}
    \caption{QUBO method (a) The solving graph is computed and (b,c) embedded onto the QPU respecting the constraints by considering edges as off diagonal terms of the QUBO matrix (d) The atom is impinged with a detuned laser with a given amplitude during some range of time. (f) The experiment is measured and samples are retrieved. The MIS represent the possible tuples between a satellite, a time step, and a request.}
    \label{fig:qubo_illustration}
\end{figure}
\subsubsection{Implementation}
For the implementation, we follow this python package~\url{https://github.com/pasqal-io/qubo-solver}.

\section{SPOT library development}
\label{sec:spot_library}
In order to reproduce the results, one can find the python package written during the projects at~\url{https://github.com/thalesgroup/spot}.
Before running into numerical experiments, we present our library developed along our work in order be as close to operational constraints and close to the hardware at hand.

\subsection{Satellite environment}
The satellite position can be obtained from a TLE string. It is propagated with the OREKIT python package. We allow the user to define a satellite from a TLE string or by reading files that are used in production.

The satellite is considered as a Reinforcement Learning environment, which means that it has a STEP method. This allows to propagate the satellite from one time step to another.
It can compute data take opportunities given a set of requests on Earth which are defined as the longitude and latitude in a separate file.

\subsection{Maximum Independent Set Solvers}
The base class of the MIS solver allows to take satellites, requests, priorities, and generate solving graphs on which we need to solve the MIS to retrieve a relyable plan for the next day.

There exist several child classes that implement the solvers to extract the MIS.
\subsubsection{Networkx}
First, the quick method contained in the NETWORKX python graph management package is provided, and can run to obtain base lines for our experiments.
\subsubsection{Rydberg}
Second, the Rydberg solvers implement routines to connect to the QPU, or emulate them locally.
Both the autoencoder and the QUBO methods are supported.

\section{Numerical experiments}
\label{sec:numerical_experiments}

\subsubsection{Visiting cities}
In order to simplify the problem for the numerical experiments, we choose $N_S$ virtual satellites that correspond to the trajectories of the ISS on $N_S$ different days.
These virtual satellites will have to schedule an observation of a set of $N_O$ cities during one day. 
Figure~\ref{fig:visiting_cities} illustrates the trajectory of three virtual satellites during one day along with the target visiting cities.

\subsubsection{Benchmark}
In order to benchmark our strategies, we design a dataset with random cities and fake satellites trajectories.
The impact of the number of satellites with the MIS strategy has already been covered by~\cite{eddy2021maximum}, so we fix the number of satellite to 3.
The number of requests per day varies from 10 to 2000.
We also vary the discretization time step, which is a hyper parameter of the algorithm proposed in the present work.
The goal is to answer the following question: is there a region where we could expect a speedup from QPU based on neutral atoms.
\begin{figure}[htpb!]
    \centering
    \includegraphics[width=\linewidth]{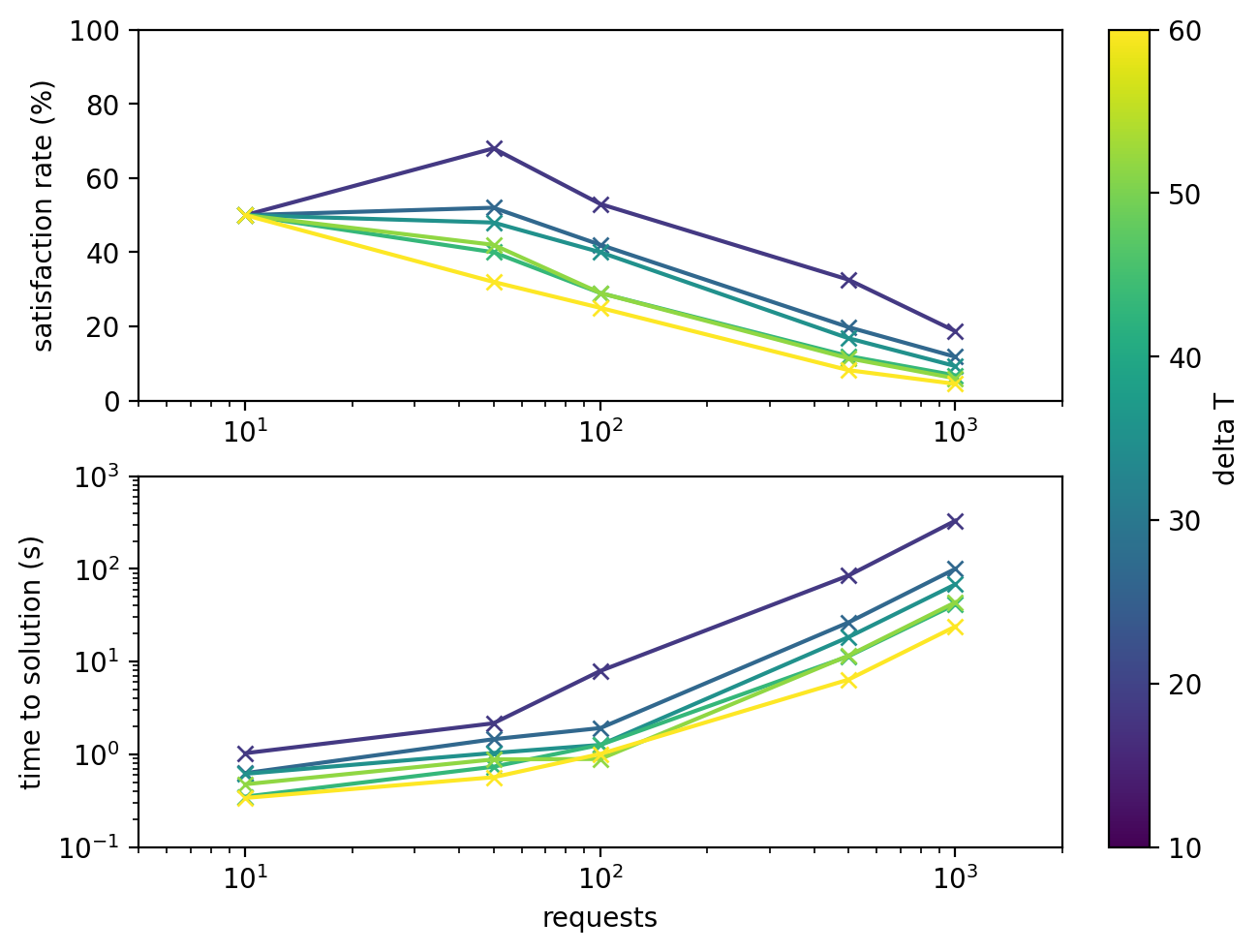}
    \caption{Benchmark: classical. Top figure: global satisfaction rate. It represents the number of requests planned divided by the total number of requests that were expected. As the number of requests increases, the problems gets more difficult, and the satisfaction rate drops. Also, if the time step duration is decreased, the completion rate improves. Bottom: global time to solution. The smaller the graphs, the faster: the smallest graphs are obtained with a large time step duration.}
    \label{fig:benchmark_results_edges_method}
\end{figure}
Figure~\ref{fig:benchmark_results_edges_method} shows the satisfaction rate achieved with the MIS networkx and QUBO solvers in the first row.
We observe that the global satisfaction rate decreases when increasing the discretization time step.
It also decreases when the number of requests increases.

The global times to solution in seconds are reported in the second row.
The spike exponentially with the number fo requests, but decrease when using a coarse discretizing time step, which is expected as there are less nodes with $\Delta T=60s$ than with $\Delta T=10s$.

In order to understand more precisely how the algorithm works, we now explore the subgraphs more thoroughly.

\begin{figure}[h!]
    \centering
    \includegraphics[width=\linewidth]{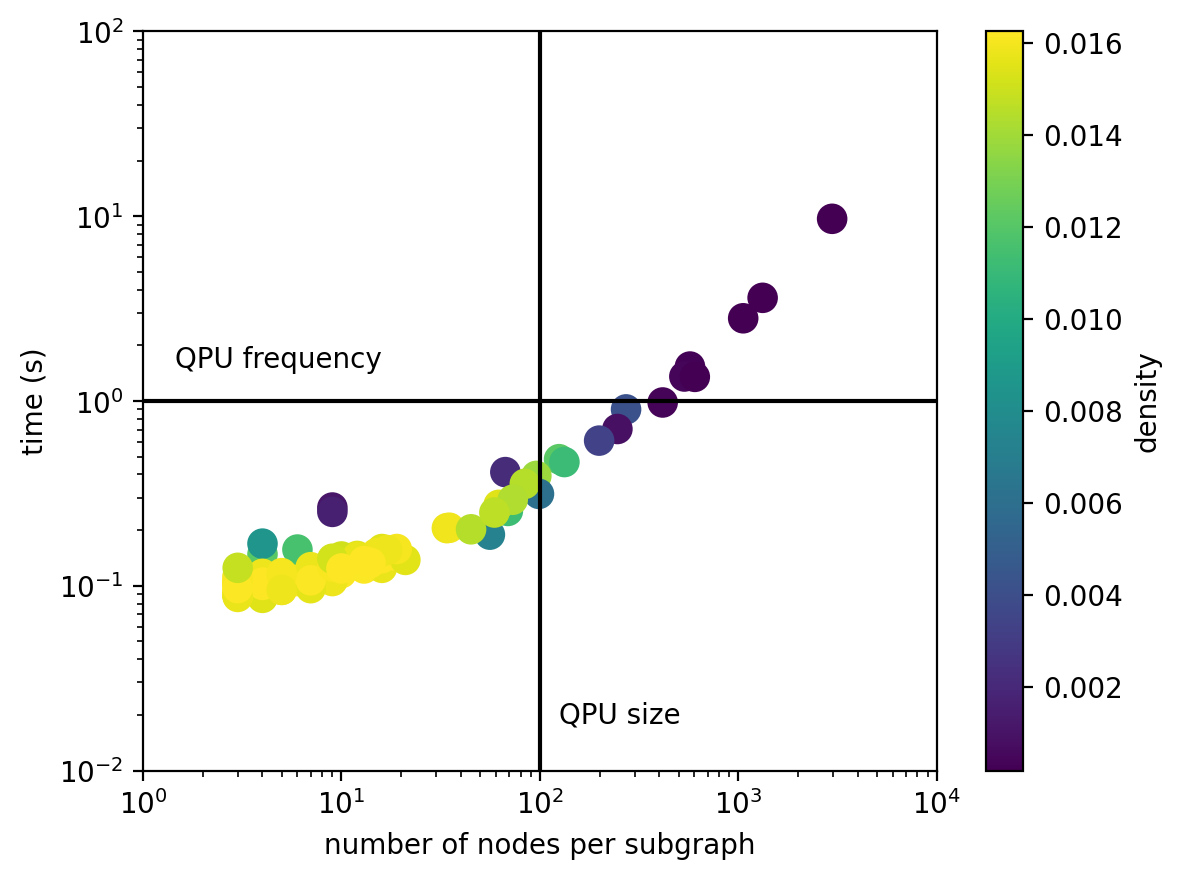}
    \caption{Time vs subgraph size: edges method. The colors represent the density. We observe that to fit on the QPU, the number of nodes for each subgraph should be less than 100. If it is higher, it uses the DECOMPOSE routine developed in the QUBOSolver. Also, if the time to solution is less than 1/1Hz (the QPU sampling rate), we shall not expect any speedup. However, QPU are getting bigger and experiments already show that all the benchmark could be leveraged by state of the art QPUs on research devices.}
    \label{fig:time_size_edges}
\end{figure}

\begin{figure}[h!]
    \centering
    \includegraphics[width=\linewidth]{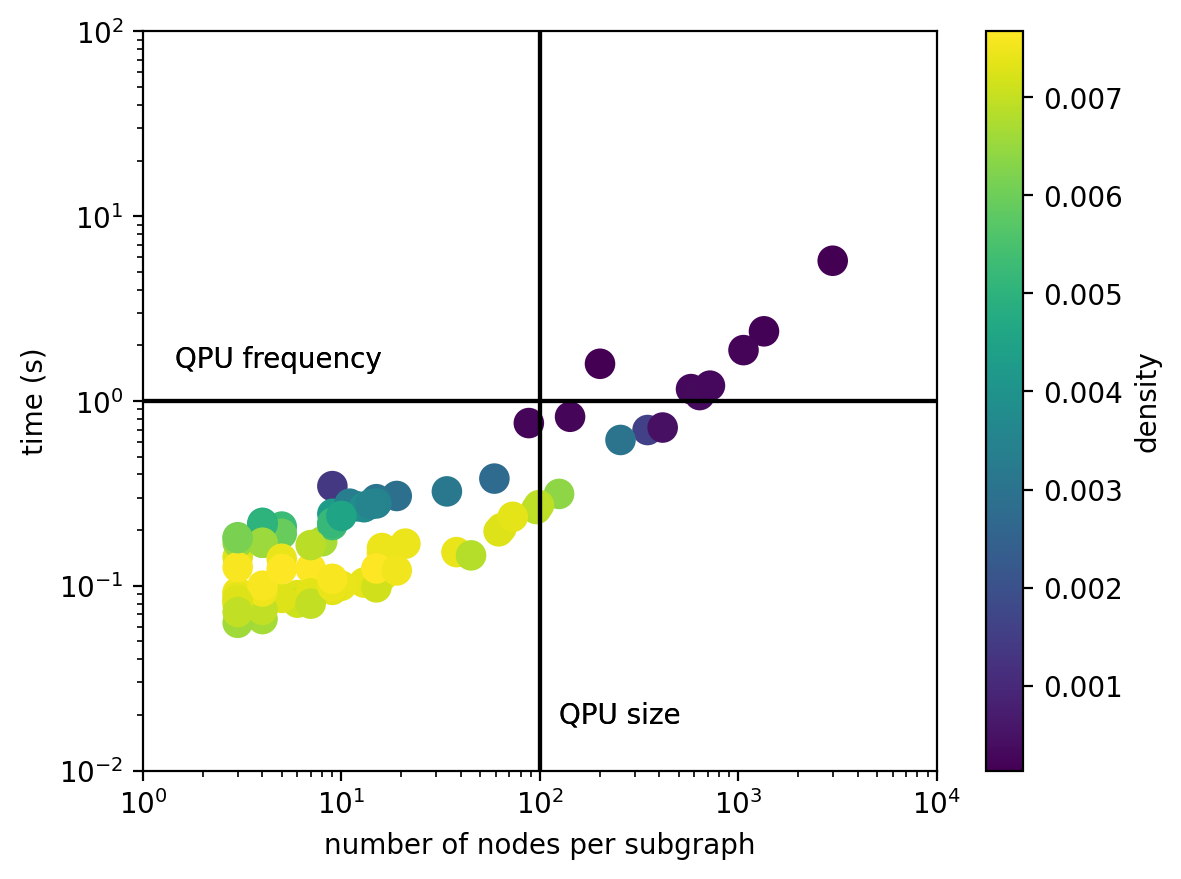}
    \caption{Time vs subgraph size: agile method}
    \label{fig:time_size_agile}
\end{figure}

With these figures in hand, it is hard to extrapolate on the potential advantage of using a QPU to solve the problem.
In order to better see when a QPU could be useful, we project the time to solution for each subgraphs versus the number of nodes per graph.
Figure~\ref{fig:time_size_edges} and~\ref{fig:time_size_agile} show the resulting figure with a density plot.
We note that most points are concentrated in the region where there are less than 100 nodes, which means that these graphs are potentially candidates for being solved on the QPU at our disposal which hosts up to a hundred of nodes.
\begin{figure*}[htpb!]
    \centering
    \includegraphics[width=0.33\linewidth]{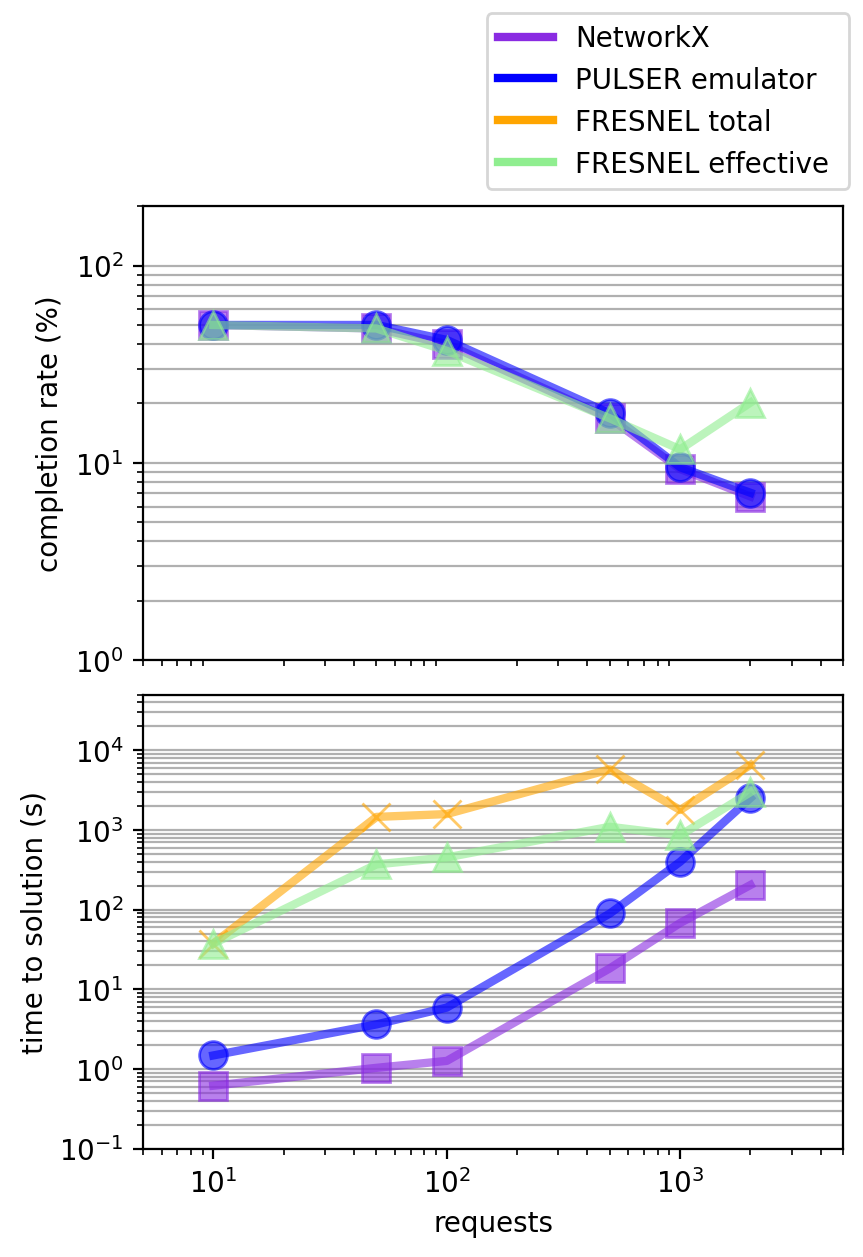}
    \includegraphics[width=0.33\linewidth]{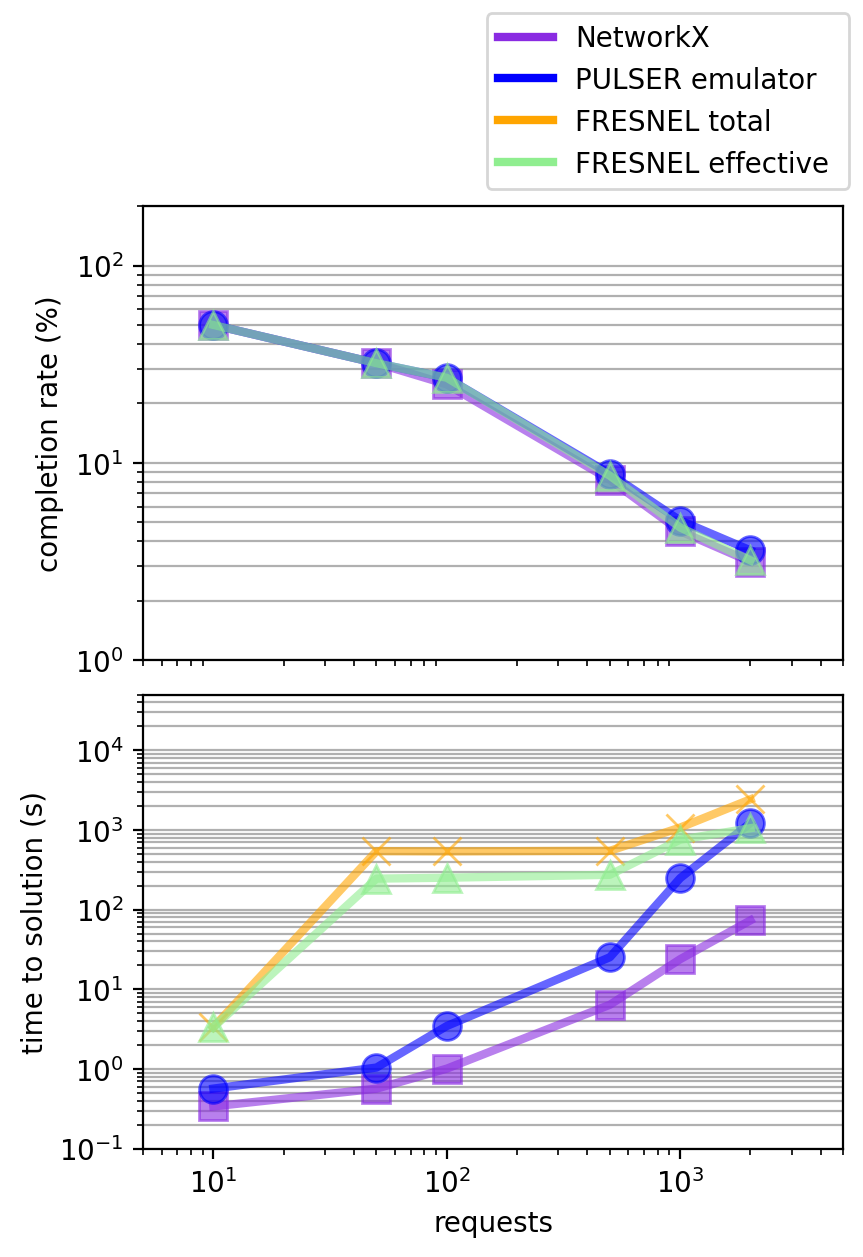}
    \includegraphics[width=0.33\linewidth]{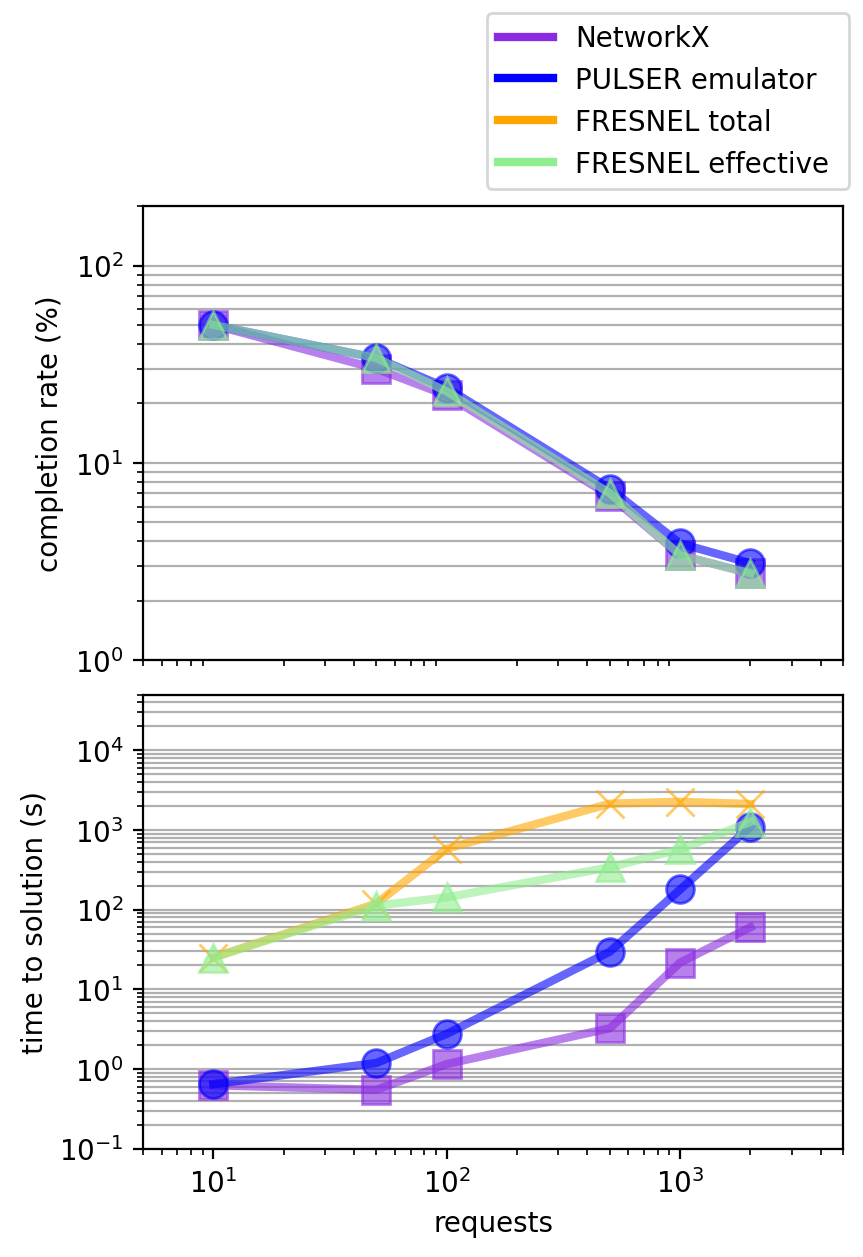}
    \caption{Performance metric/completion rate (top figures) and time to solution (bottom figures) against number of requests for a scenario with 3 satellites. Classical reference is drawn in violet. Blue curve represent the local QPU emulator without noise and with QUBOSolver. Green represent effective QUBOSolver run. Orange is QUBOSolver with waiting time. Left: $\Delta T=30s$. Middle: $\Delta T=60s$. Right: $\Delta T=90s$
    }
    \label{fig:final_results}
\end{figure*}
Figure~\ref{fig:final_results} shows the final benchmark results.
Our findings are the following:
\begin{enumerate}
    \item For the completion rates, all runs converge to a similar solution up to 1000 requests.
    \item There is one outlier for the QUBOSolver experiment with $\Delta T=30s$. 
    \item The total time to solution is prohibitive as there was only one QPU available at the time of the simulations. But hopefully we were able to retrieve the effective run time in green, which are computed as total run time minus waiting time for QPU availability.
    \item The effective run time increases with the number of requests because of the DECOMPOSE method used to split the graphs into smaller ones.
    \item The slope of time to solution vs number of requests QPU solver is lower than the one for the emulations or classical solver. We can expect that this will remain true for large numbers of requests.
    \item Even though we do not report any speedup for now, we expect the future QPU generations to have a frequency of 10 orders of magnitudes higher, which could mean that in a near future we could expect to recover classical time to solution or maybe better ones.
\end{enumerate}

\section{Conclusion}
\label{sec:conclusion}
The goal of the work was to establish the potential utility of using a QPU based on Rydberg atoms to solve the Satellite Mission Planning problem.
A QPU of 100 atoms was at disposal.
In order to achieve our goals, we developed a python package which is available at~\url{https://github.com/thalesgroup/spot}.
The package allows to treat satellite trajectories along a day.
Each satellite scans some region of Earth and is able to take observations on with an optical instrument.
Our strategies rely on formulating the problem into a Maximum Independent Set problem on a graph, which we can solve with two different type of solvers on Rydberg architectures.
This graph can be decomposed into unconnected components.
We focused our efforts on the QUBO solver because finding the optimal embedding of the subgraphs was the fastest method.

Also, because some of the subgraphs could not fit into an atomic register (namely more than a hundred of atoms), we decided to rely on a decomposition method developed by Pasqal.
We extracted as much statistics as we could from the resulting subgraphs and highlighted the fact that the Satellite Mission Planning problem might benefit from current or futur QPUs based on the neutral Rydberg atoms architecture.
It appears that the sampling rate of the QPU (less than a Hz) does not depend on the size of the graph.
Having this order of magnitude in mind, we compared this figure to the time taken by a classical solver implemented in the NetworkX python package.
In order to validate the proposed approach, we tried to emulate the hardware at hand with already available libraries.
We resorted to a decomposition strategy to obtain subgraphs that fit on the QPU, which is essential to treat big instances.
We run the experiments for the EDGES method for three discretizing time steps (30s, 60s and 90s) which are the ones that should be as close as possible as the AGILE method.
The results show that we recover the main performance metric (completion rate) of the classical solver (NetworkX solver).
The time to solution are reported, both total (including waiting on the jobs' queue) and effective (total minus the waiting time).
We report that the time to solution is comparable for the emulator and the real runs on QPU for at least 1000 requests. However, the time to solution is still higher than the classical one. This needs to be put in perspective with the fact that future QPU sampling frequency will be enhanced up to a factor 10 for each shot per subgraph, which means that we could expect speedup compared to classical naive MIS algorithms.

\begin{acknowledgments}
The authors thanks AQUAPS funding project by GENCI 2021-2024 and QR4EO ESA 2025 - 2026.
\end{acknowledgments}

\bibliographystyle{unsrt}
\bibliography{main}

\end{document}